# Variational Quantum Solutions to the Advection-Diffusion Equation for Applications in Fluid Dynamics

Reuben Demirdjian[1], Daniel Gunlycke[2], Carolyn A. Reynolds[3], James D. Doyle[3], Sergio Tafur[4]

*[1]National Research Council, Monterey, CA, 93943, USA*

*[2]Naval Research Laboratory, Washington, DC, 20375, USA*

*[3]Naval Research Laboratory, Monterey, CA, 93943, USA*

*[4]NavAir Warfare Centers –Aircraft Division, Patuxent River, MD, 20770, USA*

*Corresponding author*: Reuben Demirdjian, Reuben.Demirdjian.ctr@nrlmry.navy.mil

## Abstract

Constraints in power consumption and computational power limit the skill of operational numerical weather prediction by classical computing methods. Quantum computing could potentially address both of these challenges. Herein, we present one method to perform fluid dynamics calculations that takes advantage of quantum computing. This hybrid quantum-classical method, which combines several algorithms, scales logarithmically with the dimension of the vector space and quadratically with the number of nonzero terms in the linear combination of unitary operators that specifies the linear operator describing the system of interest. As a demonstration, we apply our method to solve the advection-diffusion equation for a small system using IBM quantum computers. We find that reliable solutions of the equation can be obtained on even the noisy quantum computers available today. This and other methods that exploit quantum computers could replace some of our traditional methods in numerical weather prediction as quantum hardware continues to improve.

# 1  Introduction

Numerical Weather Prediction (NWP) is a subset of the broader field of fluid dynamics that seeks to provide solutions to systems of partial differential equations (PDEs), namely the Navier-Stokes equations, ultimately amounting to solving an initial value problem. One challenge in NWP is that the spatial scales for a forecast can range from molecular to planetary scale ($10^7$ m), sometimes called the "tyranny of scales" [1], resulting in a continual drive towards creating operational weather models with finer and finer grid spacing. Finer grids in turn require larger calculations with increased power demands. With current model designs, downscaling an operational model from 10 km to 1 km horizontal resolution would increase the power consumption by approximately three orders of magnitude [2]. Furthermore, "resolutions of 1 – 5 km … are crucial for resolving convection" and with current model designs, a "high-performance computer of unprecedented dimension and cost would be required" to perform operational simulations at those resolutions [2]. This suggests that operational NWP is approaching a power consumption limit that would halt resolution-based forecasting skill improvements [3].

Quantum Computing (QC) has the potential to alleviate the power consumption roadblock faced in NWP. For example, the estimated energy consumption to perform Google's recent quantum supremacy calculation on a quantum computer was recently compared with that of Summit, the world's most powerful supercomputer [4, 5]. It was found that the energy consumptions of Summit and the quantum computer were 780,000 kWh and 1.4 kWh, respectively [4]. There are two main reasons for the 5.7 orders of magnitude ratio: (i) the calculation on the quantum computer took far less time to run, 200 seconds vs 2.5 days, and (ii) the power consumption of the single-chip quantum computer is far less than that of Summit, which has roughly the size of two tennis courts [5].

Another challenge that NWP faces is the end of Moore's law [6] — the observation that computing power doubles approximately every 18 months due to the increase in transistor chip density. Transistor size cannot shrink indefinitely and while there is uncertainty on when the limit will be reached, some studies suggest that could be as soon as 2030 [7]. Regardless of the precise timing, Moore's law coming to an end will directly impact operational NWP. Capping computational power will impede forecast skill improvements by limiting several modelling facets such as parameterization methods, data assimilation, and further downscaling of model resolution.

The challenges posed by the increasing power consumption and the end of Moore's law do not necessarily imply the end of forecast skill improvements, as there are several other promising approaches that could improve upon current model design. Some potential methodologies include improved parameterization theories, new model frameworks like machine learning [8-10], mathematically solving the governing equations of fluid dynamics [11], or, as focused on in this study, the exploitation of QC. [12]

Recently, there has been a flurry of publications exploring the possibility to efficiently solve partial differential equations, which are directly relevant for NWP, using QC [12-28]. Some of these algorithms are developed for long-term applications relying on fault-tolerant QC, others for nearer-term applications using Noisy Intermediate Scale Quantum (NISQ) computers. Although the number of gate operations in NISQ algorithms are limited owing to quantum decoherence in NISQ computers, they can still offer a computational advantages over their classical counterparts. A prominent example is the variational style algorithm [29] that uses a hybrid quantum-classical approach to leverage the advantages both have to offer. The present study applies a variational approach to investigate what is required of NISQ algorithms to offer a computational advantage for NWP applications.

## 2 Methods
### 2.1 The Advection-Diffusion Equation

The advection-diffusion equation, a generalization of the Burgers equation, is a non-linear PDE that describes the change in a quantity undergoing advection and diffusion. In the case where the quantity of interest is momentum, the one-dimensional form of this equation is given by

$$\frac{\partial u}{\partial t} = -u\frac{\partial u}{\partial x} + \nu\frac{\partial^2 u}{\partial x^2}, \quad (1)$$

where $u$ is velocity, $t$ is time, $x$ is the spatial coordinate, and $\nu$ is the diffusion coefficient, herein set to $5 \times 10^{-2}$. This equation is an example of a fundamental initial value problem in fluid dynamics. While it does not have any direct application in operational NWP, solving this simplified equation using QC as a proof-of-concept is an initial step toward solving equations with increased complexity.

### 2.2 Advection-Diffusion Matrix Equation for Quantum Systems

A quantum algorithm is applied that solves the initial value problem for dissipative quadratic PDEs through use of the forward Euler method to set up a linear system of equations [20]. This specific class of problems includes several equations relevant in fluid dynamics, e.g., the advection-diffusion equation and Navier-Stokes equations. One issue in solving non-linear equations using QC is the fact that quantum mechanics is fundamentally a linear theory. To overcome this challenge, a linearization method is performed that introduces powers of the variables in the nonlinear differential equation and maps the problem to an infinite sequence of coupled linear differential equations [20]. An approximation of the system of equations is made by truncating the infinite series, whereby including more terms will increase the accuracy of solution.

The vector space dimension $N$ of the linear system of equations for this method is given by

$$N = N_T n \frac{n^\tau - 1}{n - 1}, \quad (2)$$

where $n$ is the number of nonlinear differential equations, $\tau$ is the truncation order, and $N_T$ is the number of discrete times, including zero (cf. Ref. [20]). Although the dimension $N$ grows exponentially with $\tau$, the accuracy of the solution also "converges" exponentially with $\tau$ for the advection-diffusion equation, allowing us to choose $\tau$ small, e.g., $\tau \leq 5$ [20]. Since a low truncation order can be safely used, the most important variable for the size of $N$ is $n$, which in NWP is the number of prognostic variables multiplied by the number of horizontal and vertical grid cells.

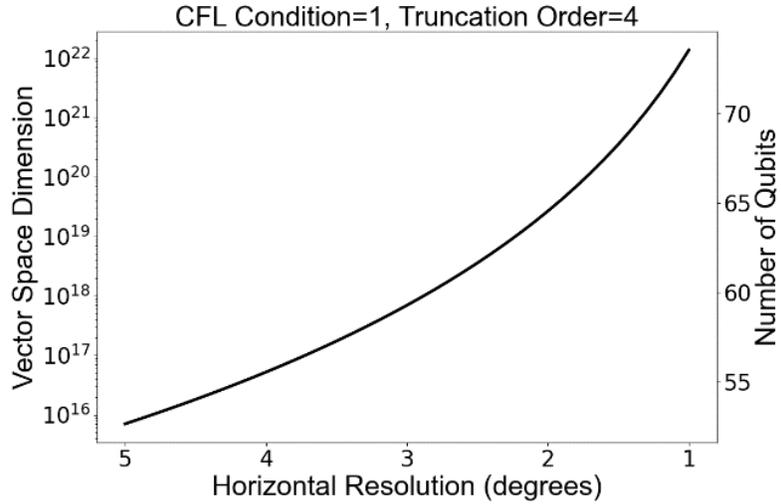

**Fig. 1** The vector space dimension required to solve the advection-diffusion equation using the methods discussed in this paper as a function of the horizontal resolution of a two-dimensional 10-day global forecasting model. The right y-axis shows the number of qubits required to process a corresponding Hilbert space of equal or greater dimension on a quantum computer.

A rough estimate is performed using (2) to determine if solving a global NWP model using QC is a feasible prospect. Consider the model representing a 10-day global forecast with a 5-degree horizontal resolution, 1 vertical level, and 5 prognostic variables. Assuming that $u_{max} = 100$ m/s and the Courant-Friedrichs-Lewy (CFL) condition is set to 1 for numeric stability, then $\Delta t = 5574$ s, which amounts to $N_T = 156$ total times. From (2) it follows that this requires a vector space dimension on the order of $10^{15}$. Moreover, assuming an efficient state preparation method exists, then a quantum computer would require 52 low-noise qubits to build and solve this linear system of equations, a prospect that is within the realm of possibility within a decade [30]. Note that this vector space dimension is not the same used in a conventional NWP model on a classical computer to solve this type of problem because the quantum algorithm requires a linearization of the terms resulting in additional variables [20]. Figure 1 repeats this calculation by varying the horizontal resolution but keeping the CFL condition constant. While the vector space dimension ranges from $10^{15}$ to $10^{22}$, the corresponding number of required qubits only ranges from 52 to 73. This illustrates that exponentially large problems could in principle be solved using only a linear amount of QC resources. Furthermore, it suggests that the vector space dimension of the linear system of equations is not the primary obstacle for solving PDEs on a QC. A discussion about other obstacles that must be overcome in order to solve large fluid dynamics problems is provided in section 4.

### 2.3 Matrix Decomposition

To be able to perform QC operations, we must first map the computational vector space to the Hilbert space of a quantum register inside a quantum computer. As this quantum register comprises qubits, the Hilbert space is a tensor power of two-dimensional Hilbert spaces of the qubits in the register. Furthermore, we require that all the states of these Hilbert spaces are unit normalized, which implies that all QC operations must be unitary. This requirement is automatically met if we decompose the linear operator describing the system of interest as

$$A = \sum_{l=1}^{L} c_l A_l, \qquad (3)$$

where $\alpha_l$ is a complex coefficient, $A_l$ is a unitary operator, and $L$ is the number of nonzero coefficients in the decomposition. Each unitary operator $A_l$ is given by

$$A_l = \otimes_q \sigma_{lq}, \qquad (4)$$

where each $\sigma_{lq}$ is either the two-dimensional identity operator $I_q$ or one of the Pauli operators $X_q$, $Y_q$, and $Z_q$ acting on the states of qubit $q$ in the register. For an $N$x$N$ matrix, the decomposition can be performed as a preprocessing step involving between $O(N)$ and $O(N^2 \log_2 N)$ operations depending on the sparsity of the operator matrix [31].

## 2.4 Quantum Solutions to Linear System of Equations

While the most advantageous linear system of equation solvers are out of reach for NISQ hardware, the Variational Quantum Linear Solver (VQLS) algorithm can theoretically provide an advantage over classical computers for large systems when solving the $A|x\rangle = |b\rangle$ problem [32]. This algorithm uses the popular quantum-classical variational hybrid method [29] to solve for $|x\rangle$ leveraging the fact that a quantum computer can efficiently determine the expectation value of some observable while the classical computer can use this expectation value to calculate a cost function. In practice, we first prepare a random initial parameter vector $\boldsymbol{\theta} = \boldsymbol{\theta}^{init}$, where $\boldsymbol{\theta} = (\theta_0, \theta_1, \ldots, \theta_P)$ consists of $P$ components of rotation angles between 0 and $2\pi$. The vector $\boldsymbol{\theta}$ specifies the quantum state $|x(\boldsymbol{\theta})\rangle = V(\boldsymbol{\theta})|0\rangle$, where $V(\boldsymbol{\theta})$ describes the quantum circuit (or ansatz) and $|0\rangle$ is the initially prepared quantum state on the register Hilbert space. The algorithm then iterates over the parameter vector $\boldsymbol{\theta}$, which is updated to minimize a cost function $C(\boldsymbol{\theta})$ that describes how "close" $A|x(\boldsymbol{\theta})\rangle$ is to $|b\rangle$. Optimization of the cost function occurs when $|x(\boldsymbol{\theta})\rangle \approx |x\rangle$ to within some termination condition.

The local cost function discussed in [32] and applied by both [33, 34] is used, which avoids the barren plateau issue discussed in [35], and converges on the global minima regardless of the starting point. This is given by

$$C(\boldsymbol{\theta}) = \frac{\langle x(\boldsymbol{\theta})|H|x(\boldsymbol{\theta})\rangle}{\langle \psi(\boldsymbol{\theta})|\psi(\boldsymbol{\theta})\rangle}, \qquad (5)$$

where $|\psi(\boldsymbol{\theta})\rangle = A|x(\boldsymbol{\theta})\rangle$ and

$$H = A^\dagger U \left( I - \frac{1}{Q} \sum_{q=1}^{Q} |0_q\rangle\langle 0_q| \otimes I_{\hat{q}} \right) U^\dagger A, \qquad (6)$$

where the unitary operator $U$ prepares $|b\rangle$ such that $U|0\rangle = |b\rangle$, $Q$ is the number of qubits in the register, and $I_{\hat{q}}$ is the identity operator for all qubits except $q$. Inserting Eq. (3) and (6) in Eq. (5), and using the identity $|0_q\rangle\langle 0_q| = (I_q + Z_q)/2$ yields

$$C(\boldsymbol{\theta}) = \frac{1}{2} - \frac{1}{2Q}\left(\frac{\sum_{q=1}^{Q}\sum_{ll'} c_l c_{l'}^* \delta_{ll'}^q}{\sum_{ll'} c_l c_{l'}^* \beta_{ll'}}\right), \quad (7)$$

where

$$\delta_{ll'}^q = \langle x(\boldsymbol{\theta})|A_{l'}^\dagger U Z_q U^\dagger A_l|x(\boldsymbol{\theta})\rangle, \quad (8)$$

and

$$\beta_{ll'} = \langle x(\boldsymbol{\theta})|A_{l'}^\dagger A_l|x(\boldsymbol{\theta})\rangle. \quad (9)$$

Herein, we use a modified version of the ansatz circuit 9 from [36] for $V(\boldsymbol{\theta})$, as shown in Figure 2a. This ansatz has been selected to constrain the solution space to include only real solutions. The circuit $U$ that prepares $|b\rangle$ is shown in Figure 2b where the angle $\phi$ is determined from the initial condition of $u(x, t = 0)$ discussed in section 3.

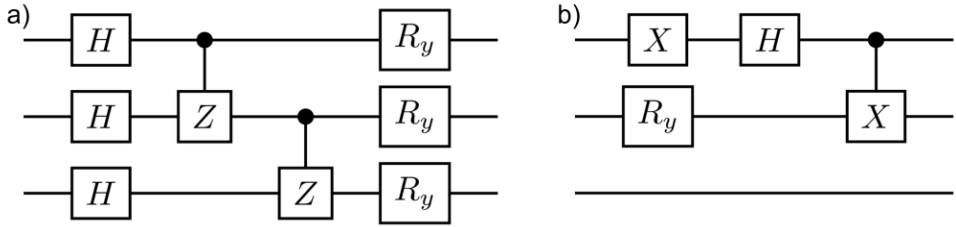

**Fig. 2** a) A unit of the ansatz that is repeated in the full quantum circuit depending on the specified circuit depth. H is the Hadamard gate, Z is the Pauli Z-gate, and $R_y$ is the rotation gate about the y-axis for angle $\theta$ discussed in text. b) The quantum circuit used to prepare the b-vector. X is the Pauli X-gate, and the angle $\phi$ is used to rotate $R_y$.

One key difference in the implementation of the VQLS method herein versus that in Ref. [32] is that the present study does not use the Hadamard test or Hadamard-overlap test methods to estimate the expectation values. While those methods have the advantage of only needing to measure the ancillary qubit, they also have the disadvantage of requiring that all unitary matrices from the linear combination of unitaries method be controlled to the ancillary qubit, thereby adding substantial two-qubit gate noise. Instead, the QISKIT [34] Pauli expectation method is used that converts the operator into Pauli strings, diagonalizes the ansatz circuit in each basis of the Pauli strings, and finally measures every qubit. The advantage of this method is that it reduces the required number of two-qubit gates. The consequence is that each qubit must be measured individually. This method is appropriate for small circuits but future implementations of this algorithm should consider both the 2-qubit gate noise and readout assignment error when deciding between the discussed methods.

There are several optimizers that could be applied to implement the VQLS method. Herein, we use the Stochastic Perturbation Simultaneous Approximation (SPSA) method [37, 38]. This particular optimizer requires fewer function evaluations to estimate the gradient than other methods, which saves valuable quantum computing resources. The chosen hyper-parameters, in the notation of Refs. [37] and [38], are: $\alpha = 0.602, \gamma = 0.101, A = 10, a = 4, c = 0.1$. Convergence is determined when 5 successive iterations have a tolerance below $2 \times 10^{-2}$. The SPSA optimizer is initialized with random starting locations within the domain during each run of the VQLS algorithm. The convergences of the 24 runs performed in this study are shown in Figure 3 where rapid convergence towards zero is observed in the first 40 iterations with slower convergence thereafter.

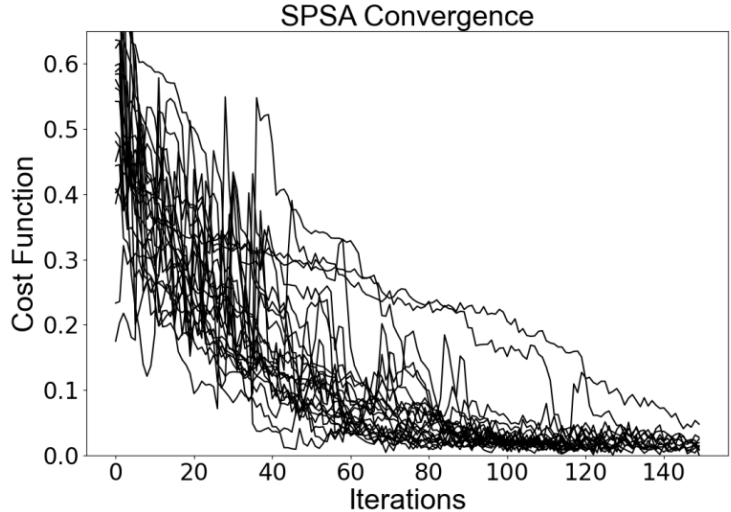

**Fig. 3** The local cost function as a function of the iteration using the SPSA optimization method for the 24 real QC runs.

### 2.5 IBMQ Hardware

The VQLS algorithm was run on three different IBM-Quantum [34] systems: (i) Cairo v1.0.2 – a 27 qubit system with a quantum volume [39] of 64 using the Falcon r5.11 processor, (ii) Hanoi v1.0.2 – a 27 qubit system with a quantum volume of 64 using the Falcon r5.11 processor, and (iii) Montreal v1.10.5 – a 27 qubit system with a quantum volume of 128 using the Falcon r4 processor. Table 1 lists an instance of the calibration data around the time the calculations herein were performed.

**Table 1** An instance of calibration data of the three IBM quantum computers used in this study obtained at the time the calculations herein were performed.

|  | Pauli-X | Average CNOT | Readout | T1 (ms) | T2 (ms) | Readout Length (ns) |
|---|---|---|---|---|---|---|
| Cairo | 3e-4 | 1e-2 | 1.3e-2 | 90 | 100 | 730 |
| Hanoi | 2e-4 | 1e-2 | 1.5e-2 | 130 | 120 | 750 |
| Montreal | 5e-4 | 1e-2 | 3e-2 | 100 | 75 | 5000 |

## 3 Results

The one-dimensional advection-diffusion equation is discretized to a stencil with $n = 4$ and initial conditions $u(x, t = 0) = \sin(\kappa x)$ where $x \in [0, L_D]$, $\kappa = 2\pi/L_D$, and $L_D = 1$ is the spatial domain size. With these initial conditions, the analytic solution to Eq. (1) is given by $u(x, t) = e^{-\kappa \nu t} \sin(\kappa x)$. Following the method in Ref. [20], the operator $A$ in the linear system of equations $A|x\rangle = |b\rangle$ for $N_T = 3$ is represented by a matrix of the form

$$A = \begin{pmatrix} I & & \\ -(I + M\Delta t) & I & \\ & -(I + M\Delta t) & I \end{pmatrix}, \quad (10)$$

where $I$ is a unit matrix, $M$ is a matrix describing the system, and $\Delta t$ is a time step size. Linearizing our system, ie. letting $\tau = 1$, we find that

$$M = \begin{pmatrix} -0.9 & 0.45 & 0 & 0.45 \\ 0.45 & -0.9 & 0.45 & 0 \\ 0 & 0.45 & -0.9 & 0.45 \\ 0.45 & 0 & 0.45 & -0.9 \end{pmatrix}, \quad (11)$$

and $\Delta t = 0.25\ s$.

Since the first block row of $A$ contains only the identity, it is straightforward to reduce the dimension of our computational vector space from $N = N_T n = 12$ to $N = 8$. This reduction transforms the state $|b\rangle$ to the form $|b\rangle = \frac{1}{\sqrt{2}}\left(\cos\frac{\phi}{2}, -\sin\frac{\phi}{2}, \sin\frac{\phi}{2}, -\cos\frac{\phi}{2}, 0, 0, 0, 0\right)^T$ where $\phi = -160.725$ follows from our initial conditions. Our reduced description allows us to bijectively map the computational space to the Hilbert space of a quantum register containing $Q = 3$ qubits, as the dimension of this Hilbert space is $2^Q = 8$, ie. the same as the computational space.

To demonstrate a proof of concept, a 24-member ensemble of the quantum solution for the advection-diffusion equation is produced on IBM's Cairo, Hanoi, and Montreal machines using 3 of their 27 available qubits and implementing the maximum 8,192 shots for each circuit. Four units of the circuit in Fig. 2a are used to form the ansatz. Each quantum state generated by this ansatz is uniquely specified by a parameter vector comprising 12 components, one for each $R_y$ gate. The cost function per iteration is shown in Figure 3 for the 24 simulations. Most trajectories are qualitatively similar and approach a cost function value approximately $10^{-2}$.

Three different types of solutions to the advection-diffusion equation are shown in Figure 4: (i) the analytic solution described in section 2.1, (ii) a solution to the matrix equation $Ax = b$ computed classically, and (iii) the solution to $A|x\rangle = |b\rangle$ obtained using the VQLS method. Note that if the quantum solution (iii) were perfect, it would reproduce the classical solution in (ii) and not the analytic solution in (i), as the linear system of equations are themselves only approximate solutions to (1). However, a comparison of Figs. 4a and 4b shows that even for the truncation order $N = 1$ [20] the linear system of equations produce adequate solutions for our purposes. Qualitative comparisons of the average of all 24 quantum solutions (bold lines in Fig. 4c) and the classical solutions (Fig. 4b) clearly demonstrate that reasonable solutions can be obtained using the VQLS method. The root mean square error of the 24-solution average at $t = 0.25$ s is 0.010 m/s and at $t = 0.5$ s is 0.021 m/s, corresponding to 6% and 15% relative errors, respectively.

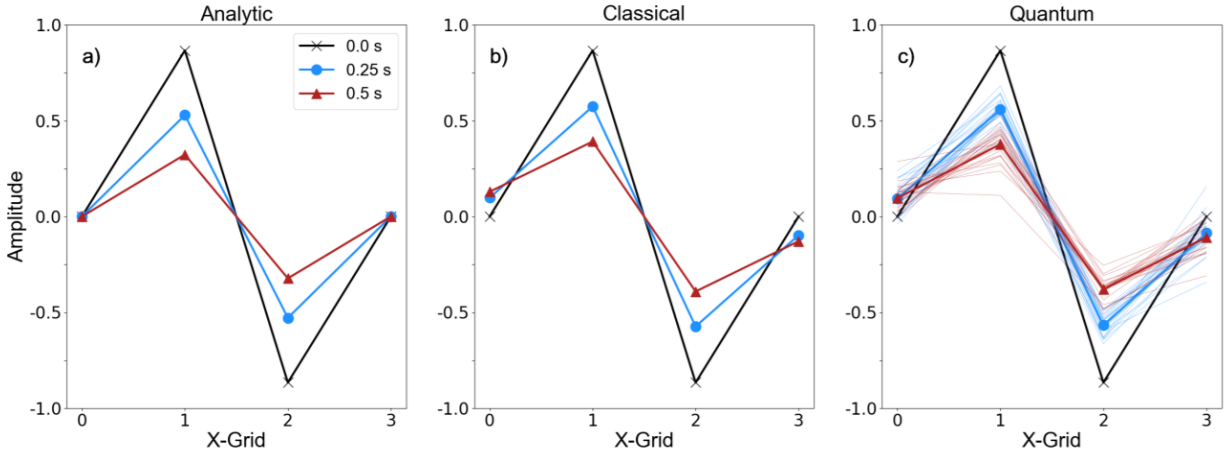

**Fig. 4** Solutions to the advection-diffusion equation at three time steps for the a) analytic solution, b) classical solution of the linear system of equations discussed in the text, c) same as (b) except for the quantum solutions of the 24 different runs on IBM's quantum hardware (thin lines) and average of the 24 runs (thick lines). For all plots the initial condition is shown in black, $0.25\ s$ time step in blue, and $0.5\ s$ time step in red.

While the mean quantum solution (Fig. 4c) looks nearly identical to that of the classical solution (Fig. 4b), there is a clear spread in the individual solutions. It is therefore logical to ask: do the individual solutions contain biases as the optimizer iterations tend toward infinity? In other words, what is the sensitivity of the solution accuracy to the stopping iteration? To investigate this, an arbitrary run is selected where the solution at each grid point is shown in Supplemental Figures 1 and 2 for $t = 0.25$ s and $t = 0.5$ s, respectively. The supplemental figures illustrate that only 1 out of the 8 solutions do *not* have an inherent bias after convergence while 7 of 8 do exhibit some kind of bias. This implies that the sign of the errors in each of the solutions may not be randomly based on which iteration the solution happens to be stopped on, but rather that after reasonable convergence the sign of the error is already determined. This might necessitate multiple integrations for an unbiased estimate.

## 4  Discussion and Conclusions

This study is focused on combining several novel algorithms to (i) set up a non-linear partial differential equation into a linear system of equations [20], (ii) perform a Pauli decomposition on the resulting matrix equation [31], and (iii) solve the decomposed linear system of equations on IBM's quantum hardware using the VQLS method [32]. While this investigation is not the first to solve partial differential equations on real hardware [19, 21], it is the first to solve a meteorologically relevant computational problem that demonstrates the feasibility for quantum computers to become a potential tool for numerical weather prediction.

The question remains: can the methods applied in this study be scaled up to solve much larger systems of equations and thereby build more sophisticated fluid models? In some particular cases it is possible, however, this algorithm cannot be scaled up to solve arbitrary problems in its current form. The bottleneck in the workflow is the cost function in the VQLS algorithm that can require an enormous number of circuits as the number of terms in the linear combination of unitaries grows. As there are $QL^2$ and $L^2$ coefficients $\beta'_{ll}$ and $\delta^q_{ll'}$ in Eq. (8) and (9), respectively, the number of circuits required for each iteration of the VQLS is

$$N_C = (Q + 1)L^2 . \qquad (12)$$

Some symmetries are exposed in the $\beta_{ll'}$ term by observing that $\beta_{ll'} = 1$ when $l = l'$ and that $\beta_{ll'} = \beta_{l'l}$, bringing a small reduction in the number of circuits as shown in Fig. 5 [32]. Similar symmetries were found to empirically exist for the $\delta^q_{ll'}$ term that allow for an even greater reduction by observing that $\delta^q_{l'l} = \delta^q_{ll'}$, reducing the overall number of circuits by about a factor of two from the baseline case (Fig. 5). It is important to note that the latter symmetry was found empirically to work for the specific case outlined in this study and that it is not in general true. A useful point of reference for Fig. 5 is that the current maximum $N_C$ that can be submitted to IBM's quantum systems is 900. Therefore, any algorithm that requires an $N_C$ significantly greater than 900 cannot be practically implemented at this time. Even in the full symmetry case, the quadratic relationship prevents the methods discussed in this paper from being scaled up substantially since $N_c$ quickly becomes too large to be realistically run on a QC.

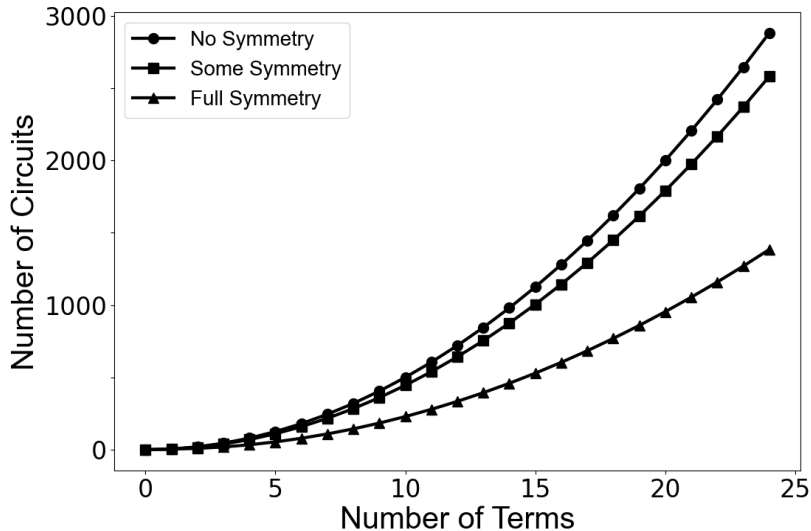

**Fig. 5** The number of circuits required as a function of the number of terms in the Pauli decomposition of the matrix A. Three cases are shown with various amounts of circuit symmetry used to reduce the total number of circuits (see text for discussion).

The specific type of case in which the VQLS can be implemented is therefore when $N_C$ is relatively small. It is unlikely that an arbitrary matrix will decompose into a small number of unitaries, so unless care is taken to prepare a specific matrix, the cost function for the VQLS algorithm will require some improvements to be useful for solving non-linear differential equations in form investigated here. A current challenge is therefore to determine if the $N_C$ scaling can be improved upon thereby opening the doors for the VQLS algorithm to be potentially useful for NWP studies on a larger scale. However, this is not a long-term challenge of fault-tolerant systems because in the future we will be utilizing completely different algorithms to solve this type of problem.

**Acknowledgements** RD, DG, CAR, and JDD has been supported by the Office of Naval Research (ONR) through the NRL Base Program, PE 0601153N. We acknowledge quantum computing resources from IBM through a collaboration with the Air Force Research Laboratory (AFRL).

## Declarations

**Conflict of interest** The authors declare that they have no conflict of interest.

**Data Availability Statement** Due to confidentiality agreements, supporting data can only be made available to bona fide researchers subject to a non-disclosure agreement. Details of the data and how to request access are available from Reuben Demirdjian at the U.S. Naval Research Laboratory.

## Supplemental Figures

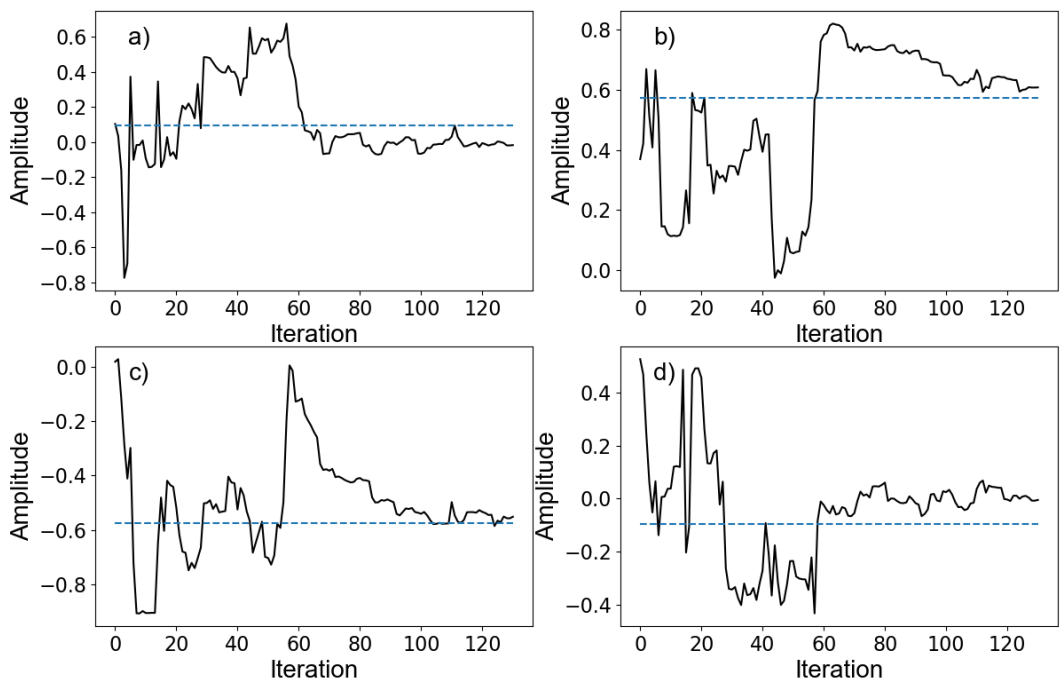

Time Step t=0.25 s

**Supplemental Figure 1** Convergence of the solution in black at the $0.25\ s$ time step as a function of iteration for a) grid point 1, b) grid point 2, c) grid point 3, and d) grid point 4. The dashed blue line is the classical solution for which the quantum solution is expected to converge onto. This particular run was performed on the IBM's Cairo machine.

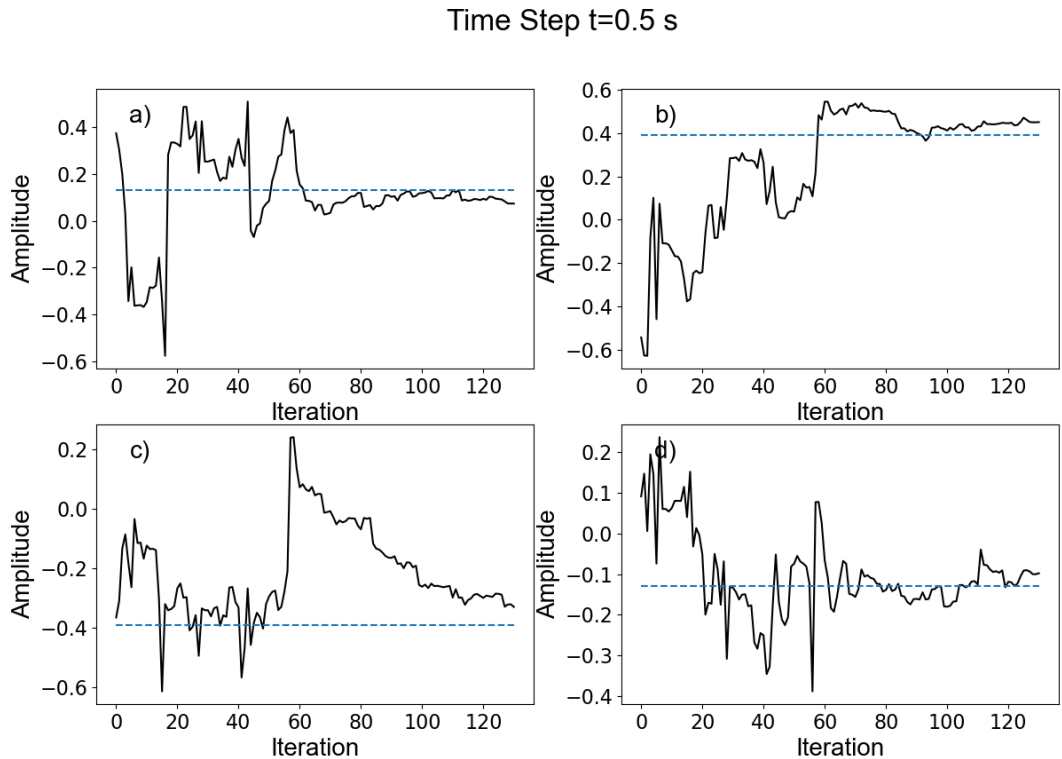

**Supplemental Figure 2** Same as Supplemental Figure 1 except for the $0.5\ s$ time step.